\documentclass[prb,onecolumn,aps,showpacs,groupedaddress]{revtex4}
\usepackage{epsfig,epsf,psfrag}
\usepackage{graphicx}
\usepackage{epic,eepic}
\usepackage{color,pstcol}
\begin{document}
\title{Detecting Majorana bound states}
\author{Colin Benjamin}
\altaffiliation{Present address: Center for Simulational Physics, Dept. of Physics \& Astronomy, Univ. of Georgia, Athens, GA 30602, USA}
\email[E-mail me at: ]{colin.nano@gmail.com}
\affiliation {Quantum Information Group, School of Physics and Astronomy, University of Leeds, Woodhouse Lane, Leeds LS2 9JT, UK.}

\author{Jiannis K. Pachos}
\affiliation {Quantum Information Group, School of Physics and Astronomy, University of Leeds, Woodhouse Lane, Leeds LS2 9JT, UK.}
\begin{abstract}
We propose a set of interferometric methods on how to detect Majorana bound states induced by a topological insulator. The existence of these states can be easily determined by the conductance oscillations as function of magnetic flux and/or electric voltage. We study the system in the presence and absence of Majorana bound states and observe strikingly different behaviors. Importantly, we show that the presence of coupled Majorana bound states can induce a persistent current in absence of any external magnetic field.
\end{abstract}
\maketitle

 \section {Introduction}

  One of the main interests in current research on quantum computation is to find new materials that facilitate the construction of large scale quantum computers. The main impediment is decoherence and the generation of errors. A promising way of reducing the effect of decoherence and errors is to employ quantum systems that have topological characteristics\cite{dassarma}. Recently, topological insulators have been considered that can be tuned to support topological states, such as Majorana bound states. There have been several proposals\cite{kitaev,dassarma} how one could encode information with Majorana fermions that is protected against a variety of errors. Majorana fermions which can also occur in highly correlated systems like $p_{x}+ip_{y}$ wave superconductors\cite{janik}, the $\nu=5/2$ fractional quantum hall state\cite{read}, at the boundary of superfluid 3He-B\cite{volovik} and finally in superconducting graphene\cite{wilczek} have the special characteristic that they are their own antiparticles. They are also predicted to appear as low energy excitations in Kitaev's two-dimensional spin-1/2 system on a honeycomb lattice\cite{kitaev,brennen,jiannis}.
\begin{figure}
\centerline{\includegraphics[scale=0.43]{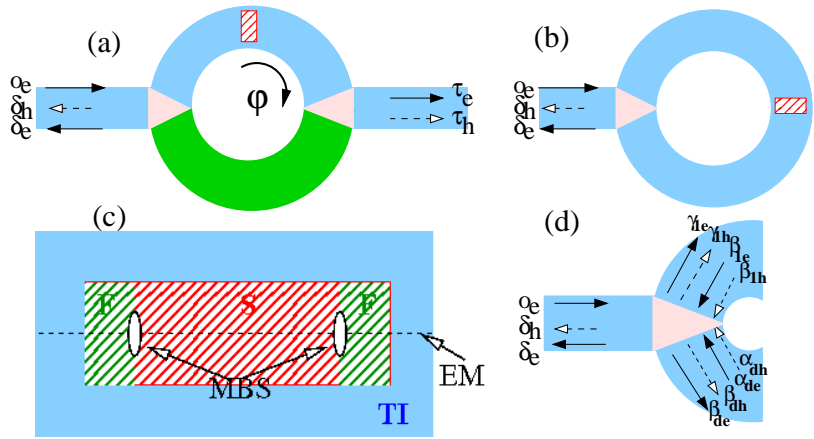}}
\caption{(Color online) An overview of the setting from the top. The 2D topological insulator (TI, in blue) is made into an Aharonov-Bohm interferometer (a). In the upper arm of the ring,  MBS (white ellipses at edge of Ferromagnet(F) and Superconductor(S)) occur as shown in (c). Topological edge modes (EM), circulate along the edges, interacting with the MBS. A magnetic flux penetrates the ring while an electric voltage covers the lower arm (in green) of the ring. (b) A loop made up of a TI which supports MBS. (d) Representation of incoming and outgoing waves from left coupler as in (a). \label{scheme}}
\end{figure}

The aim of this work is to detect Majorana bound states, at the interface between  topological insulators with superconducting and magnetic correlations, addressed also in Refs.[\onlinecite{kane,been}]. Coupled Majorana bound states (implying, two Majorana bound states which are interacting) can encode a qubit non-locally and obviate local environmental perturbations\cite{been}. Their detection, so far, has been difficult as Majorana fermions are neutral quasi-particles. To overcome that recent works proposed to employ Dirac to Majorana fermion converters\cite{akh,fu}. There, Dirac particles emitted from a source are converted to Majorana's and then reconverted back at the drain. In our work Majorana bound states  are efficiently monitored by a mesoscopic Aharonov-Bohm interferometer. In particular, the presence of Majorana bound states can be probed by the symmetry of the non-local conductance as function of the magnetic field or an applied electric field. Further, we show that the presence of coupled Majorana bound states could induce persistent currents in a topological insulator ring in absence of any magnetic flux.

\section {The proposed physical Model} To detect Majorana bound states(MBS) that exist in the topological insulator(TI)\cite{been}, we look at an Aharonov-Bohm (AB) interferometer made up of a 2D TI (e.g., $Bi_{2}Se_{3}$\cite{moore} or a HgTe quantum well\cite{expt}) as depicted in Fig.~1(a). In a TI, spin–orbit coupling causes an insulating material to acquire protected edge or surface states. A magnetic flux exists in the center of the AB interferometer. The regions of the interferometer are labeled in Fig.~1(a). In the upper arm of the ring, at the interface between s-wave superconducting(S) and a thin ferromagnetic layer(F), MBS (white ellipses at the interface) occur as shown in (c). Topological edge modes(EM), circulate at the edges, interacting with the MBS. A magnetic flux penetrates the ring while an electric voltage covers the lower arm (in green) of the ring. The superconductor and Ferromagnet deposited on top of the TI, via the proximity effect, induce superconducting and ferromagnetic correlations in the substrate. The places where these correlations intersect is where MBS occur. The ring is connected to two leads on either side. The left lead is at potential $V_1$ while $V_{2}=0$(see Fig. 2). The full Hamiltonian in the upper and lower arms of the interferometer satisfies \begin{equation}
(vp\tau_{z}\sigma_{z}+(eV-E_{F}+eA/\hbar c)\tau_{z}) \Psi = E \Psi, \label{topo-H} \end{equation}
wherein $p=-i\hbar\partial/\partial x$ is the momentum operator, $E_F$ the Fermi energy, $v$  the Fermi velocity, $eV$ the electric field applied to the lower arm only and $A$ defines the magnetic vector potential. The four component wave-function $\Psi=(\Psi_{e\uparrow},\Psi_{e\downarrow},\Psi_{h\uparrow},\Psi_{h\downarrow})^{T}$, while the $\tau$ matrices mix the e and h blocks of the Hamiltonian. The eigenstates of Hamiltonian (\ref{topo-H}), can be calculated by considering plane wave solutions. The Hamiltonian for the superconducting-magnet interface (white ellipse at edge of red and greens dashed areas as in Fig.~1(c)) is that for the MBS,
\begin{equation} H_{M}=-\sigma_{y}E_{M}\label{majo}\end{equation}. As discovered by Fu and Kane\cite{kane}, a MB state appears at the intersection of the magnet-superconductor interface with the edge of the TI. The $4\times4$ S-matrix of scattering via the MBS, Ref. [\onlinecite{been}], can be written as ${S_{Maj}}=s^{ab}_{ij}$ where $\{a,b\}=\{e,h\}$ and $\{i,j\}=\{1,2\}$. The elements are determined as follows
\begin{eqnarray}
s^{ee}_{11}&=&s^{hh}_{11}=1+ix, s^{ee}_{22}=s^{hh}_{22}=1+ix',\nonumber\\ s^{eh}_{11}&=&s^{he}_{11}=ix, s^{eh}_{22}=s^{he}_{22}=ix',\nonumber\\ s^{ee}_{21}&=&-s^{ee}_{12}=s^{eh}_{21}=-s^{eh}_{12}=y,\nonumber\\ s^{hh}_{21}&=&-s^{hh}_{12}=s^{he}_{21}=-s^{he}_{12}=y\nonumber\\ x&=&\frac{\Gamma_{1}(E+i\Gamma_{2})}{z},x'=\frac{\Gamma_{2}(E+i\Gamma_{1})}{z},y=\frac{E_{M}\sqrt{\Gamma_{1}\Gamma_{2}}}{z},\nonumber\\ z&=&E_{M}^{2}-(E+i\Gamma_{1})(E+i\Gamma_{2})
\label{maj-reason}
\end{eqnarray}
 In the above equation, E is the incident electron energy, $\Gamma_{1/2}$ are the strengths of coupling to left/right arms and $E_M$ the strength of coupling between the individual MBS.

 \section {Edge modes and scattering matrices} To understand the edge modes and the type of scattering matrix needed to describe them, we show in Fig. 2, the edge modes flowing in our system and the way MBS affect them. We start with a related system, in Fig. 2(A) a quantum hall conductor with an AB flux is shown. A localized state around the hole develops and is sensitive to the flux, the outer edge states are not\cite{buti-ti}. However the outer edge states determine the net conductance when a voltage is applied to left lead. Thus if there is no scattering between outer and inner edge states the conductance is independent of flux. For conductance to be sensitive to flux one has to couple both these states. Thus as seen in Fig. 2(B) we couple the outer edge states to the inner states via the two couplers. The couplers induce inter edge scattering as shown by dashed lines. A $3\times3$ matrix can effectively describe this scattering process as there are $3$ outgoing and $3$ incoming modes. In the TI with a hole shown in Fig. 2(C), the edge states occur in pairs but are of opposite spins (black:up, white:down), there are also two localized counter-propagating edge states which develop around the hole. If as we assume there is no scattering between states with opposite spins then up spin conductance will be equal to that for down spins. Couplers induce backscattering for all edge modes while MBS mixes electron and hole edge modes and also backscatters. We restrict ourselves only to describing electron spin up and hole spin down edge modes as shown in Fig. 2(D) this leads to two counter propagating electron and another two hole edge modes around the flux. This leads to a $6\times6$ S-matrix for either couplers while a $4\times4$ matrix couples inner edge modes. Similarly one can construct the electron spin down and hole spin up edge modes. The total conductance will be twice that for incident electron spin-up edge modes since there is no spin-flip scattering.
 \begin{figure}
\centerline{\includegraphics[scale=0.45]{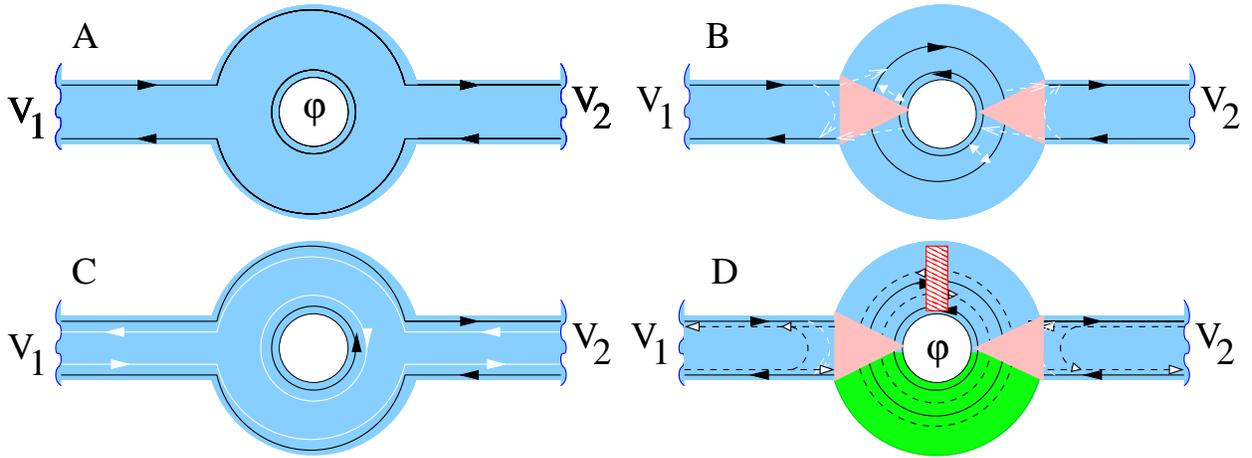}}
\caption{(Color online) Edge modes and back scattering. (A) Quantum Hall conductor with localized flux. (B) Coupling between outer and inner edge modes in a Quantum Hall conductor. (C) A TI with a hole, spin -up(black) and down(white) edge modes counter propagate along the hole. (D) A MBS scatterer (red shaded area on the upper arm of the AB ring) induces e-h mixing and back scattering.  \label{scheme1}}
\end{figure}
\begin{figure}[t] \vskip 0.36in \centerline{\includegraphics[scale=0.45]{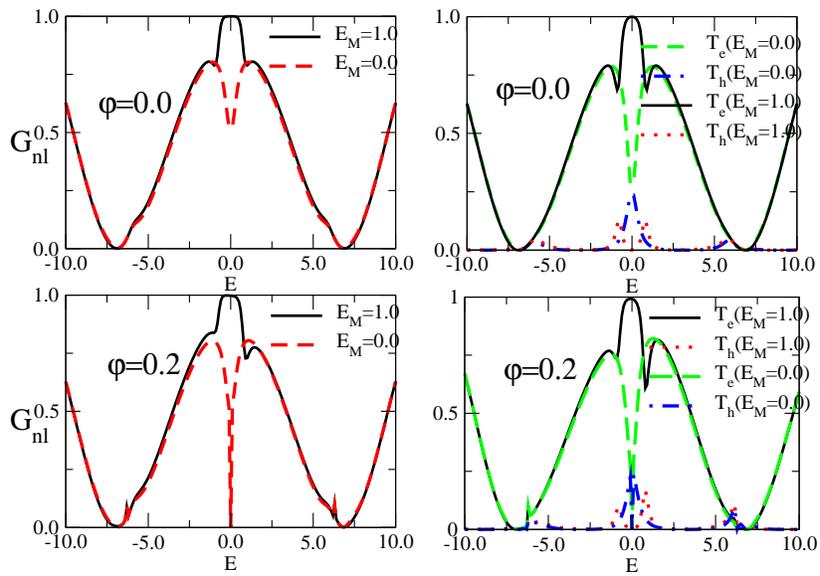}} \caption{(Color online) $G_{nl}$, in units of $e^{2}/\hbar$, as function of incident electron energy E (controlled via a gate voltage) in units of $\Delta$. The dashed and solid lines in the left panel are for individual ($E_{M}=0$) and coupled MBS ($E_{M}=1.0$), $E_M$ in units of $\Delta$ the superconducting gap.\label{nonlocal}}
\end{figure}
 In Fig.~1(a) and 2(D), the length of upper arm is $l_u$ and of lower arm is $l_d$, total circumference of loop is $L=l_u+l_d$. The Majorana scatterer further divides the upper arm, as shown in the figure $l_{u}=l_{1}+l_{2}$ with $l_{1}=l_2$. The loop is connected to two current leads on either side.  The couplers (triangles) in Fig.~1 which connect the leads and the loop are described by a scattering matrix $S$. The S matrix for the left coupler yields the amplitudes $O_{1}=(\delta_{e},\delta_{h},\gamma_{1e},\gamma_{1h},\beta_{de},\beta_{dh})$ emanating from the coupler in terms of the incident waves $I_1=(o_{e},o_{h},\alpha_{de},\alpha_{dh},\beta_{1e},\beta_{1h})$, and for the right coupler yields the amplitudes $O_{2}=(\tau_{e},\tau_{h},\gamma_{de},\gamma_{dh},\beta_{2e},\beta_{2h})$ emanating from the coupler in terms of the incident waves $I_2=(i_{e},i_{h},\xi_{de},\xi_{dh},\gamma_{2e},\gamma_{2h})$. The S-matrix for either of the couplers\cite{butipra}, left and right, is given by
 \begin{equation}
 S=\left(\begin{array}{ccc} -(a+b)\bf{I}      & \sqrt\epsilon\bf{I}&\sqrt\epsilon\bf{I}\\ \sqrt\epsilon\bf{I}& a  \bf{I}          &b   \bf{I}         \\ \sqrt\epsilon\bf{I}& b    \bf{I}        &a\bf{I} \end{array} \right)
 \end{equation}
 with $a=\frac{1}{2}(\sqrt{(1-2\epsilon)} -1), b=\frac{1}{2}(\sqrt{(1-2\epsilon)} +1)$, {\bf I} being the identity matrix. $\epsilon$ plays the role of a coupler with maximum coupling $\epsilon=\frac{1}{2}$ while for $\epsilon=0$, the coupler completely disconnects the loop from the lead. For left coupler, the waves into and out are marked in Fig.~1(d). The waves incident into the branches of the loop are related by the S matrices for left part of the upper branch by

 \begin{equation}\left(\begin{array}{c} \beta_{1e}\\ \alpha_{1e}\\ \beta_{1h}\\ \alpha_{1h} \end{array} \right) \ =\left(\begin{array}{cccc} 0     & e^{ik_e l_1} e^\frac{-i \phi l_1}{L}&0&0\\ e^{ik_e l_1}e^\frac{i \phi l_1}{L} & 0 &0&0\\ 0&0&0&e^{i k_h l_1}e^\frac{{i \phi l_1}}{L}\\ 0&0&e^{i k_h l_1} e^\frac{-i \phi l_1}{L} \end{array} \right) \left(\begin{array}{c} \gamma_{1e}\\ \xi_{1e}\\ \gamma_{1h}\\ \xi_{1h} \end{array} \right)
 \end{equation}
 while for the right part

 \begin{equation}
 \left(\begin{array}{c} \gamma_{2e}\\ \xi_{2e}\\ \gamma_{2h}\\ \xi_{2h} \end{array} \right) \ =\left(\begin{array}{cccc} 0     & e^{ik_e l_2} e^\frac{i \phi l_2}{L}&0&0\\ e^{ik_e l_2}e^\frac{-i \phi l_2}{L} & 0 &0&0\\ 0&0&0&e^{i k_h l_2}e^\frac{{-i \phi l_2}}{L}\\ 0&0&e^{i k_h l_2} e^\frac{i \phi l_2}{L}&0 \end{array} \right) \left(\begin{array}{c} \beta_{2e}\\ \alpha_{2e}\\ \beta_{2h}\\ \alpha_{2h} \end{array} \right)\end{equation}
 and  for lower branch

 \begin{equation}\left(\begin{array}{c} \alpha_{de}\\ \xi_{de}\\ \alpha_{dh}\\ \xi_{dh} \end{array} \right) \ =\left(\begin{array}{cccc} 0     & e^{ik'_e l_d} e^\frac{i \phi l_d}{L}&0&0\\ e^{ik'_e l_d}e^\frac{-i \phi l_d}{L} & 0 &0&0\\ 0&0&0&e^{i k'_h l_d}e^\frac{{-i \phi l_d}}{L}\\ 0&0&e^{i k'_h l_d} e^\frac{i \phi l_d}{L}&0 \end{array} \right) \left(\begin{array}{c} \beta_{de}\\ \gamma_{de}\\ \beta_{dh}\\ \gamma_{dh} \end{array} \right)\end{equation}
 with $k'_{e}=(E+Ef+V)/\hbar v$, $k'_{h}=(E-Ef-V)/\hbar v$  are the electron and hole wave-vectors, while $k_e$ and $k_h$ are wave-vectors with $V=0$. $\frac{\phi l_1}{L}$, $\frac{\phi l_2}{L}$ and $\frac{\phi l_d}{L}$ are the phase shifts due to flux in the upper and lower branches. Clearly, $\frac{\phi l_1}{L}+\frac{\phi l_2}{L}+\frac{\phi l_d}{L}=\frac{2\pi\Phi}{\Phi_0}$, where $\Phi$ is the flux piercing the loop and $\Phi_0$ is the flux quantum $\frac{hc}{e}$. The transmission and reflection probabilities from Fig.~1 and Eq.~(3) are given as follows: normal electron reflection $R_{e}=|\frac{\delta_e}{o_e}|^2$, non-local electron co-tunneling $T_{e}=|\frac{\tau_e}{o_e}|^2$, local Andreev reflection $R_{h}=|\frac{\delta_h}{o_e}|^2$ and non-local crossed Andreev reflection $T_{h}=|\frac{\tau_h}{o_e}|^2$ wherein $\tau,\delta$ are as depicted in Fig.~1(a). In the calculations we consider $e=\hbar=c=1$ and $\Gamma_{1}=\Gamma_{2}=\Gamma.$ For the setting as described in Fig. 1(b), the scattering matrices can be written in exactly similar fashion (see also Ref.[\onlinecite{buti-loop}]). The total persistent current density\cite{jayan} in a small interval $dE$ is then sum of the individual electronic and hole current densities calculated as $J=e v (J_{e}+J_{h})$, with $J_{e}= (|\gamma_{1e}|^{2}-|\beta_{1e}|^2)$, and $J_{h}= (|\gamma_{1h}|^{2}-|\beta_{1h}|^2)$.

\begin{figure}[t]
\vskip 0.1in
\centerline{\includegraphics[scale=0.45]{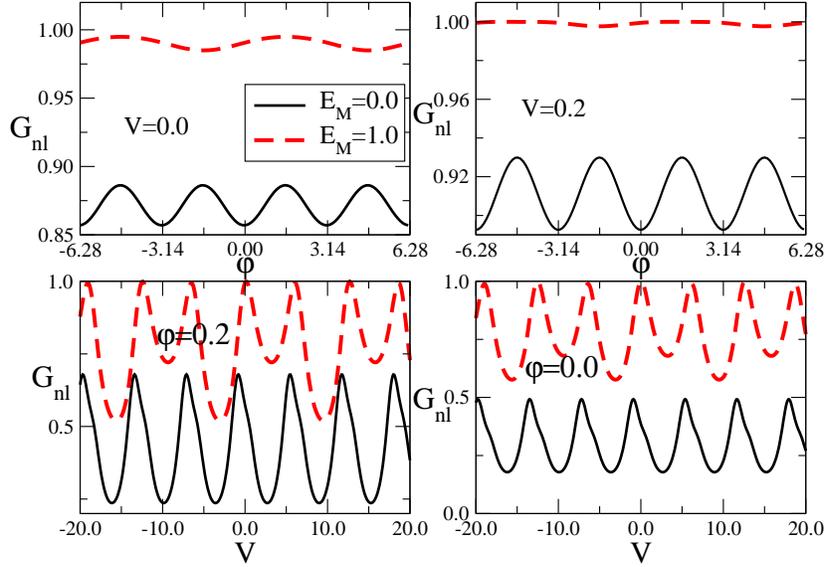}} \caption{(Color online) Non-local conductance, in units of $e^{2}/\hbar$, as function of magnetic flux (top panel) in units of $\Phi_{0}=\hbar c/e$ and electric potential (bottom panel) in units of $\Delta$ for $E=0.1\Delta$.\label{nonlocal_phi_V}} \end{figure}
\section  {Detection procedure for MBS}

To distinguish the behavior of the MBS we analytically solve Eqs.~(4-7) and derive expressions for reflection and transmission probabilities. For brevity we present plots that result from these solutions. In Figs.~3 and 4 we plot the nonlocal conductance $G_{nl}=(e^{2}/2\hbar)[1-R_{e}-R_{h}+T_{e}+T_{h}]=(e^{2}/\hbar)(T_{e}+T_{h})$. We subsumed a "$-$" sign into the hole wavevectors, indicating their opposite direction to electrons, in the S matrices. Thus in $G_{nl}$ we add the individual contributions rather than subtract. As a consistency check- the sum of probabilities $T_{e}+T_{h}+R_{e}+R_{h}=1$. Further, current conservation also holds as currents in either leads are equal $T_{e}+T_{h}=1-R_{e}-R_{h}$. The non-local conductance implies a current which appears in the right lead, while a voltage $V_1$ is applied to left lead and no voltage is applied to the right. Hence, appearance of a current is due to the non-local effect of voltage applied to the left lead.
\begin{figure}[t]
\vskip 0.39in
\centerline{\includegraphics[scale=0.45]{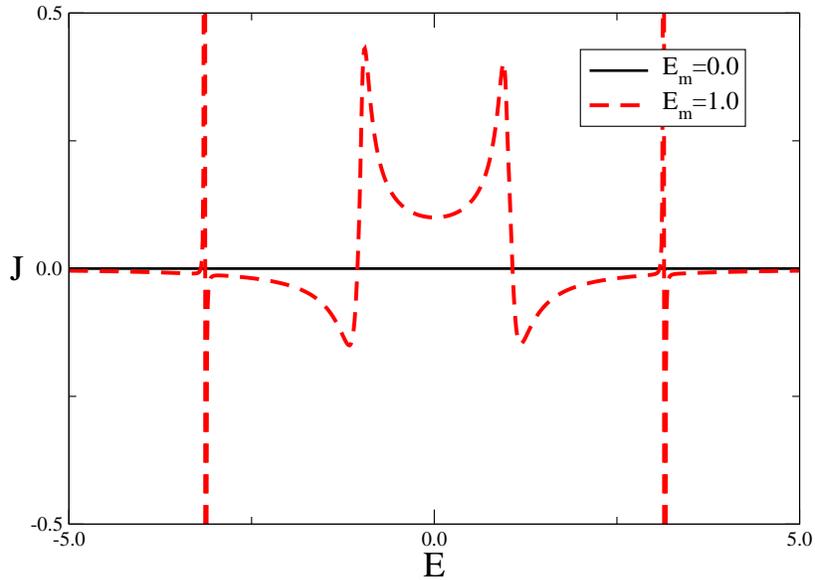}} \caption{(Color online) The current density(J) in units of $e v$ as function of the incident electron energy, for $\epsilon=1/2, \mbox{and } E_{F}=0$. A finite current (red dashed line) flows when MBS are coupled.\label{J-em}} \end{figure}
In Figs.~3-5, the dimensionless parameter $\epsilon=1/2$, Fermi energy $E_{F}=0$ and $\Gamma=0.1$ in units of $\Delta$ and lengths $l_{u}=l_{d}=1/2$ in units of $\hbar v/\Delta$. In Fig.~3, we focus on the behavior of $G_{nl}$ as function of the electronic energy (which can be tuned by a gate voltage) in the top left panel. We see there is  a pronounced dip in the case wherein MBS are decoupled. On the adjacent panel we plot the individual contributions. We see that most of the contribution to $G_{nl}$ comes from electron co-tunneling. Further, $G_{nl}(E)=G_{nl}(-E)$ for $\phi=0$, but this equality does not hold in presence of magnetic flux. Fig.~4 shows the variation of $G_{nl}$ as function of the magnetic flux. It shows $G_{nl}(\phi)=G_{nl}(-\phi)$ in the case where the MBS are decoupled, while for coupled states $G_{nl}(\phi)\neq G_{nl}(-\phi)$. The absence or presence of an electric voltage on the lower arm does not make much of a qualitative difference while a quantitative difference is manifest. On the lower panel of Fig.~4 we plot $G_{nl}$ against the electric potential $V$ on the lower arm of the ring. It is seen that while $G_{nl}(V)\neq G_{nl}(-V)$ irrespective of whether MBS are coupled or not, there is a halving of periodicity when MBS are decoupled. The reason for non-observance of magnetic field symmetry is two-fold: (i) A coupled MBS scatterer, breaks time reversal symmetry as in Eq.~(\ref{maj-reason}), e.g., $s_{12}^{ee} \neq s_{21}^{ee}$ and (ii) breakdown of Andreev reflection symmetry ($s_{11}^{eh*}(-E)=s_{11}^{he}(E)$) instead of $s_{11}^{eh*}(-E)=-s_{11}^{he}(E)$, regardless of whether MBS are coupled or not. However, for $E,\Gamma \ll \Delta$, the breakdown of Andreev reflection symmetry can be discounted as this amplitude is minimal and therefore we chose the low energy and weak coupling sectors in Figs.~3-4 so as to bring out the distinction between coupled and individual MBS in the non-local conductance versus magnetic field plots wherein breakdown of time reversal symmetry is the main reason.
\begin{figure}[t]
 \centerline{\includegraphics[scale=0.45]{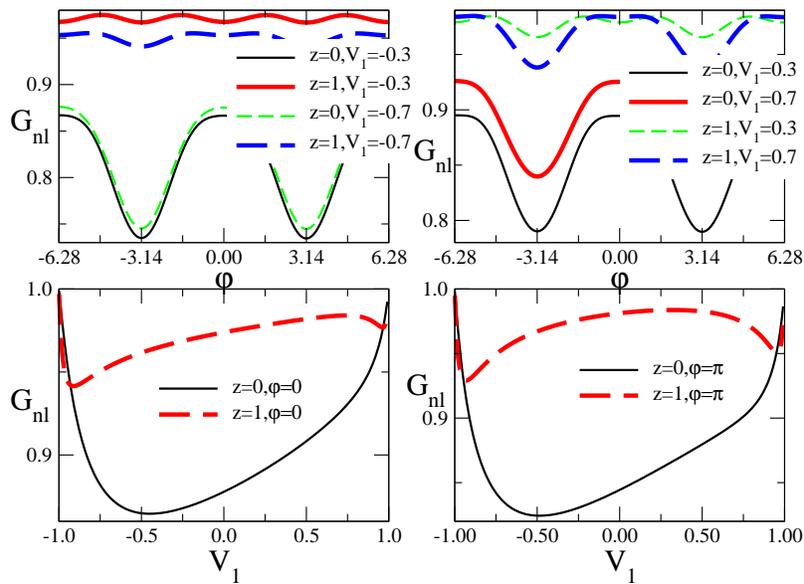}} \caption{(Color online) Non-local conductance as function of magnetic flux and voltage applied to left lead, $V_1$, for a AB ring with a s-wave superconductor in its upper arm in absence of MBS. z represents normal metal-superconductor interface strengths in units of $\Delta$.\label{nonlocal_swave}}
\end{figure}

Because, of break down of time reversal symmetry electrons and holes scattered from the coupled state get different phases when they travel from left to right and vice-versa. The same philosophy which first predicted persistent currents in mesoscopic rings due to the fact that a magnetic flux can break time reversal symmetry\cite{landauer} is once again present here. Importantly, here it is not just the magnetic flux which causes the breaking of time reversal symmetry but it is also the Majorana scattering. This implies that a circulating current can arise in the TI loop because of the scattering due to coupled MBS independent of the fact whether a magnetic flux is present or absent. To isolate this effect we calculate the persistent current for the setting described in Fig.~1(b). In presence of coupled MBS, a persistent current is induced in the TI loop while for individual (or, de-coupled) MBS such a current is absent. In Fig.~5, we plot the persistent current density (in units of $e v$), which when integrated over the energy gives us the total persistent current. Experimental detection of this persistent current would be via a measurement of the magnetic moment of the ring.

Finally, we consider the case of a similar setting (as in Fig.~1(a)) but without any MBS. This system is a normal metal AB ring with an s-wave superconductor in its upper arm. The left lead is at potential $V_1$ while no voltage is applied to the right. In Fig.~6 we plot the results for the non-local conductance $G_{nl}=(e^{2}/\hbar)[1-R_{e}+R_{h}+T_{e}-T_{h}]$. To conserve currents on either side of the superconductor one lets the potential in the superconductor float. Due to the absence of MBS the nonlocal conductance is symmetric with respect to magnetic field reversal (see top panel of Fig.~6). In the bottom panel of Fig.~6, we see $G_{nl}(V_{1})\neq G_{nl}(-V_{1})$. This is in contrast to the case where MBS are present.
\begin{table}[t]
 \begin{center}
\caption{{Detecting Majorana bound states (MBS)}}
\begin{tabular}{|c|c|c|} \hline MBS & {\em Magnetic field}& {\em Electric field/gate voltage} \\ \hline individual ($E_{m}=0$) & $G(\phi)= G(-\phi)$&  $G(E)= G(-E)$\\ \hline coupled ($E_{m}\neq 0$) & $G(\phi)\neq G(-\phi)$ & $G(E) = G(-E)$ \\ \hline Absent & $G(\phi)=G(-\phi)$ & $G(V_{1})\neq G(-V_{1}$) \\ \hline \end{tabular}
\end{center}
\end{table}

\section {Conclusions}
We have introduced a novel mechanism to detect MBS occurring in TIs. The realization and control of MBS would be the first step towards a fault tolerant quantum computer. In Table I we summarize our results. We see that coupled MBS break magnetic flux symmetry. One also sees that period doubling occurs in presence of an electric potential and in the presence of individual MBS (Fig.~3). These can be easily used as means to detect MBS. Further there is a pronounced zero-energy dip/crest in presence of individual MBS which changes into a continuous function (without a maximum or minimum) in their absence. Finally, coupled MBS, can induce a persistent current in absence of a magnetic flux in a topologically insulating loop. Although these results are derived for a loop made of a TI, the results would remain valid in case the TI is replaced with a semiconductor provided a MBS can exist in such a system\cite{sau}, since we used a generic Hamiltonian for the Majorana scatterer, see Eq.~(\ref{majo}).  An extension of the work would be to study the  shot noise generated in our settings\cite{bj}. The symmetries of this could also be a matter of interest in the process of detection.

 \section {Acknowledgments}
 The authors thank Andrej Mesaros(Leiden U.) and Stefanos Papanikolaou(Cornell U.) for useful comments on a previous version of this manuscript. This work was supported by the EU grants EMALI and SCALA, EPSRC and the Royal Society.

\end{document}